\documentclass[twocolumn,showpacs,preprintnumbers,amsmath,amssymb]{revtex4}
\usepackage{graphicx}
\usepackage{dcolumn}
\usepackage{bm}
\def\bea {\begin{eqnarray}}
\def\eea {\end{eqnarray}}

\def\be {\begin{equation}}
\def\ee {\end{equation}}

\begin{document}
\title{van der Waals hadron resonance gas and QCD phase diagram}

\author{\it Nachiketa Sarkar}
\author{\it Premomoy Ghosh}
\email{prem@vecc.gov.in}
\address {Variable Energy Cyclotron Centre, HBNI, 1/AF Bidhan Nagar, Kolkata 700 064, India}   
     
\begin{abstract}
Taking into account the recently developed van der Waals (VDW) like equation of state (EoS) for grand canonical ensemble of fermions, the temperature dependent profiles of 
normalized entropy density ($s /T^3$) and the ratio of shear viscosity and entropy density ($\eta/ s$) for hadron resonance gas have been evaluated. The VDW parameters,
corresponding to interactions between (anti)baryons, have been obtained by contrasting lattice EoS for QCD matter at finite chemical potentials ($\mu_{B}$) and for $T \le$ 160 
MeV. The temperature and chemical potential dependent study of $s /T^3$ and $\eta /s$ for hadron gas, by signalling onsets of first order phase transition and crossover 
in the hadronic phase of QCD matter, helps in understanding the QCD phase diagram in the ($T, \mu_{B}$) - plane. An estimation of probable location of critical point matches
predictions from other recent studies. \\ 
\end{abstract}
\maketitle
\section{ Introduction}
Quark-Gluon Plasma (QGP) \cite {ref01, ref02}, the de-confined partonic phase of strongly interacting matter, is created in laboratories in ultra-relativistic heavy-ion collisions 
at Relativistic Heavy Ion Collider (RHIC) \cite {ref03, ref04, ref05, ref06} and Large Hadron Collider (LHC) \cite {ref07}. The QGP, evolving through a confined hadronic phase, 
eventually results into free-streaming final state particles. Lattice simulations  \cite{ref08} of quantum chromodynamics (QCD) at vanishing baryon chemical potential ($\mu_{B}$),
corresponding to top-RHIC and LHC energies, provide reliable equation of state (EoS) for both the phases of strongly interacting matter, partonic and hadronic. The hadronic 
phase of the QCD matter \cite {ref01, ref02} at zero $\mu_{B}$ can be described successfully also by the Hadron Resonance Gas (HRG) Model \cite {ref09, ref10, ref11}. The 
QCD matter at non-zero $\mu_{B}$, like the ones created \cite{ref12} in heavy-ion collisions at comparatively lower centre-of-mass energies ($\sqrt s_{NN}$) in the Beam Energy 
Scan (BES) program at RHIC, however, is less understood.\\

According to the present understanding of the QCD phase diagram in ($T, \mu_{B}$)-plane, at vanishing $\mu_{B}$ and at $T$, higher than that at a critical point, the changes 
between partonic and hadronic phases occur through a crossover \cite {ref13, ref14}. On the other side of the critical point, along the phase boundary, first order phase transition 
\cite {ref15, ref16, ref17} takes place. RHIC has collected $AuAu$ collision data at $\sqrt s_{NN}$ = 7.7, 11.5, 19.6, 27, 39, 62.4, 130 and 200 GeV, which cover a wide range 
of baryon chemical potential from $\mu_{B} \approx$ 420 to 20 MeV. Being encouraged with the non-monotonic trends in BES-data \cite{ref12}, in particular in the event-by-event 
net-proton fluctuations, the BES-II \cite {ref18} program has been planned to collect high statistics precision-data in the energy range $\sqrt s_{NN}$ = 7.7 to 20 GeV. The fixed-target
mode $Au + Au$ collisions at BES-II will cover $\sqrt s_{NN}$  = 3 GeV ($\mu_{B}$ = 720 MeV) to 7.7 GeV. Another fixed-target
experiment, the Compressed Baryonic Matter (CBM) \cite {ref19} at Facility for Antiproton and Ion Research (FAIR) will cover $Au+Au$ collision energy range of $\sqrt s_{NN}$ = 2.5 
to 4.7 GeV, corresponding to $\mu_{B}$ = 800 - 500 MeV. Also, experiments at Nuclotron - based Ion Collider fAcility (NICA) \cite {ref20} and J-PARC-HI at the Japanese proton
synchrotron accelerator facility \cite {ref21} will have heavy-ion collisions, creating high baryon density ($\mu_{B} \sim$ 850 MeV) QCD matter. While all these experiments aim to 
study the QCD phase boundary and to search for the possible QCD critical point in the non-zero, high $\mu_{B}$ range, theory supplement is not adequate yet, as reliable EoS for
strongly interacting matter at high $\mu_{B}$ is still not possible directly from LQCD formulation. The EoS for strongly interacting matter at non-zero, small $\mu_{B}$, however, is
obtained in lattice regularities from truncated Taylor expansion of thermodynamic potential and recently such an EoS has been obtained \cite {ref22} for $\mu_{B}/T$ = 1.0, 2.0 and 
2.5, the range that has been covered by the experiments in the BES program of RHIC. At this stage, in absence of $\it {ab}$ $\it{initio}$ calculations for the baryon-rich QCD matter, 
one can obtain EoS for the hadronic phase of QCD matter at small, non-zero $\mu_{B}$ in a suitable HRG model by contrasting the lattice results and, can extend the study further 
in the ($T, \mu_{B}$)-plane.\\

In this article, we present our study in terms of temperature dependent normalized entropy density, $s /T^3$ and the ratio of shear viscosity and entropy density, $\eta /s$ for hadron 
gas in a HRG model, incorporated with van der Waals (VDW) form of equation of state that was appropriately developed \cite {ref23, ref24, ref25} for grand canonical ensemble of 
fermions. By fixing the parameters of VDW EoS for fermions with the properties of the nuclear matter ground state, the location of critical point at the end of first-order nuclear liquid-gas
phase transition has been predicted \cite {ref24}. Inclusion of the VDW EoS in HRG, with the same interaction constants, gives qualitatively better results \cite {ref26}, resembling 
close to the LQCD results, in the crossover region at zero $\mu_{B}$. Estimation of QCD critical point in VDWHRG model has been attempted \cite {ref27}, by comparing the LQCD 
EoS at zero $\mu_{B}$. To study the hadronic phase of QCD matter over a wide range of $\mu_{B}$, we prefer to obtain the interaction constants by comparing lattice EoS \cite{ref22} 
for QCD matter at finite $\mu_{B}$. \\

It is important to note that the nuclear liquid-gas phase transition does not involve change in degrees of freedom and so, the signals for phase transitions and for the critical 
end point could be clearly brought out \cite {ref24} by the van der Waals EoS. On the contrary, the QCD phase transition or crossover entails change in degrees of freedom between
hadronic and partonic and so, full description of the QCD phase diagram including both the phases cannot be studied with only the hadron gas. Nevertheless, by studying temperature
dependent $s /T^3$ and $\eta /s$ of hadron gas at varied $\mu_{B}$, with a van der Waals form of EoS, contrasted with lattice EoS for QCD matter, the regions of the onset of phase
transition or crossover can be identified and thus, the region of probable QCD critical point can be estimated.\\ 

\section{Observables}
In-depth knowledge of thermodynamic variables and transport coefficients of hadronic phase is necessary for a better understanding of the QCD phase diagram. In particular, the 
entropy density that gives information about the stability of an equilibrated system and the shear viscosity, measuring the ability of a fluid medium to relax towards equilibrium after 
a shear perturbation, are the mostly studied observables in characterising the strongly interacting matter, produced in relativistic nuclear collisions. Study of $s /T^3$ and $\eta /s$
as a function of temperature can reveal signal for phase transition and crossover. While $s /T^3$ is discontinuous at a point of first order phase transition, a smooth but rapid rise 
of $s / T^3$ over a small change in $T$ signals the crossover. Similarly, a first-order phase transition presents a discontinuity in $\eta /s$ and the crossover exhibits a smooth arrival 
of $\eta /s$ at a minimum \cite {ref28, ref29, ref30, ref31, ref32, ref33}. Relevant literature provide straight forward procedures for estimation of entropy density of hadron gas, 
characterizing a thermalized hadronic phase of QCD matter. Conversely, in spite of increased activities in the study of $\eta /s$ on the basis of the prediction \cite {ref28} that the 
minimum of $\eta /s$ should lie at the critical temperature or near phase transition or rapid crossover temperature between the two phases of the QCD matter and that the conjectured
universal minimum ($\le 1/4\pi$), the KSS bound \cite {ref34}, of $\eta /s$ is valid for the QCD matter also, estimation of $\eta$ of a hadron gas is still under development stage. 
This is evident from the widely varied values of $\eta /s$ of hadron gas obtained from different model calculations.\\ 

The temperature dependence of $\eta /s$ for hadron gas, as studied  \cite {ref35, ref36, ref37, ref38, ref39} in HRG models with short range repulsive interactions introduced by 
considering finite sizes of constituent hadrons, exhibit monotonic behaviour. In general, the temperature dependent $\eta /s$, estimated in the HRG model based calculations in 
molecular kinetic theory \cite {ref35, ref36, ref37, ref38} involves a rather simple, analytical formula of the shear viscosity of a gaseous system, proportional to the number density, 
the mean free path and the average momentum of the gas molecules. At low temperature, this analytical formula yields large value of $\eta /s$, compared to that obtained from 
elaborate theoretical calculations including Kubo formalism \cite {ref40} and Chapman-Enskog (CE) approach \cite {ref41}, which are in good agreement \cite {ref42} with each 
other. The transport coefficient of hadron gas is studied using the Relaxation Time Approximation (RTA) \cite {ref44, ref45}, also. The temperature dependence of $\eta /s$ for 
hadronic matter at zero $\mu_{B}$, estimated in microscopic transport calculations \cite {ref46, ref47} with UrQMD and (conceptually similar) SMASH, employing Kubo formalism, 
result in considerably different values of $\eta /s$ at low temperature. The variance in the values of $\eta /s$ from various transport models can be attributed \cite {ref47} to varied
microscopic details that can be translated very differently into macroscopic effects. In reference \cite {ref43}, the temperature dependence of $\eta /s$ of a multicomponent hadronic 
resonance gas have been estimated by calculating $\eta$ in the CE approach and entropy density in relativistic virial expansion method using the K-matrix parametrization of hadronic
cross sections. The calculation yields half the value of $\eta /s$, reported in reference~\cite {ref46}. In reference \cite {ref45}, the minimum of $\eta /s$, following RTA method, reaches 
the crossover temperature at 245 MeV. In a hydrodynamics-based transport model study \cite {ref48}, the estimated $\eta /s$ near $T$ = 160 MeV, comes one-forth the value 
obtained in reference~\cite{ref46}. An extrapolation to the temperature dependent $\eta /s$ for QCD hadronic phase, calculated \cite {ref31} using chiral perturbation theory and the
linearized Boltzmann equation, reaches the KSS bound at $T$ $\sim$ 200 MeV. It is important to note that none of the estimations of $\eta /s$ for hadron gas at  $T$ $\approx$ 160
MeV reaches the KSS value, corroborating the conclusion \cite {ref46} that the expected range of low values of $\eta /s$ ($\sim$ 0.08 to 0.24) \cite {ref49} at RHIC might be attributed 
to the partonic phase of the QCD matter and not to the hadronic phase. \\

In constraining the regions of phase transition and crossover in QCD phase diagram with a hadronic gas model, it is, therefore, important that a study of temperature dependence 
of $\eta /s$ is complemented with a study of temperature dependence of $s /T^3$.\\ 

\section{Hadron Resonance Gas (HRG)}
\subsection{Ideal and Excluded Volume Models}
The basic version of the HRG model, formulated with the experimentally measured discrete mass spectrum of hadrons and resonance states provided in mass tables 
by particle data groups (PDGs), successfully reproduces several thermodynamic observations from LQCD calculations of strongly interacting hadronic matter \cite {ref50, ref51, 
ref52, ref53} at vanishing $\mu_{B}$. While the inclusion of all known resonances effectively takes care of the attractive interactions between the hadrons, one needs to take into 
consideration the repulsive interaction between the constituent hadrons. This is usually done in the so-called excluded volume (EV) \cite {ref54, ref55, ref56, ref57, ref58, ref59} 
model of HRG (EVHRG) by introducing the effects of Van der Waals type hadron repulsion at short distances, implemented through finite hard-core radius of constituent hadrons 
of the system. \\

The grand canonical partition function of ideal (noninteracting) hadron resonance gas are written as\cite {ref10}: 	
\begin{equation}
\ln Z\textsuperscript{id}=\sum_{i=1} \ln Z_{i}^{id} 
\label{eq:1}
\end{equation}
where $ \ln Z_{i}^{id}$ is the partition function of the $i^{th}$ particle and is given by: 
\begin{equation}  
\label{eq:2}
\ln Z_{i}\textsuperscript{id}=\pm\frac{V g_i}{2\pi^2 }\int_{0}  ^\infty p^2 dp  \ln \left\{1\pm \exp[-(E_i-\mu_i)/T] \right\},
\end{equation}
where $V$ is the volume of the system, $g_{i} $ is the degeneracy, $T$ is the temperature. $E_{i}=\sqrt{p^2+m^2_{i}}$ is the single-particle energy, $m_{i} $ is the 
mass and $\mu_{i} = B_i\mu_B+S_i\mu_s+Q_i\mu_Q $ is the chemical potential, $B_{i}, S_{i}, Q_{i}$ are the baryon number, strangeness, and charge of the particle, 
respectively, $\mu's$  are corresponding chemical potentials. The $(+)$ and $(-)$ sign corresponds to fermions and bosons respectively.\\

The thermodynamic variables: pressure $P^{id}(T,\mu)$, particle density $n^{id}(T,\mu)$, energy density $\epsilon^{id}(T,\mu)$ and entropy density $s^{id}(T,\mu)$ for the 
ideal hadron resonance gas can be written as:\\
\begin{widetext}
\begin{equation}
P^{id}(T,\mu) = \pm \sum_i\frac{g_iT}{2\pi^2 } \int_{0}^\infty p^2dp\ln \left\{1\pm \exp[-(E_i-\mu_i)/T] \right\}
\end{equation}
\begin{equation}
n^{id}(T,\mu) = \sum_i \frac{g_i}{2\pi^2 }\int_{0}^\infty \frac {p^2 dp}{ \exp[(E_i-\mu_i)/T] \pm 1}
\end{equation}
\begin{equation}
\epsilon^{id}(T,\mu) = \sum_i \frac{g_i}{2\pi^2 }\int_{0}  ^\infty \frac {p^2 dp}{\exp[(E_i-\mu_i)/T] \pm 1 }E_i
\end{equation}
\begin{equation}
s^{id}(T,\mu) =  \pm\sum_i \frac{g_i}{2\pi^2 }\int_{0}^\infty  {p^2 dp}[\ln \left\{1\pm \exp[-(E_i-\mu_i)/T] \right\} \pm \frac{(E_i-\mu_i)}{T(\exp[(E_i-\mu_i)/T] \pm 1)}]
\end{equation}
\end{widetext}

In a thermodynamically consistent Excluded Volume (EV) HRG model, as proposed by Ref \cite{ref56}, the pressure is given by: 
\begin{equation} \label{eq:16}
P_{EV}^{id}(T, \mu_1,\mu_2,...)=\sum P_i^{id}(T, \bar\mu_1,\bar\mu_2,...)
\end{equation}

The chemical potential of the $i^{th}$ particle is given by:  
\begin{equation} \label{eq:17}
\bar\mu_i=\mu_i-V_{ev,i}P_{EV}^{id}(T, \mu_1,\mu_2,...)
\end{equation}
where $V_{ev,i}=\frac{16}{3} \pi r_i^3$ excluded volume of the $i^{th}$ hadron  with hard core radius $r_{i}$		 	 			 		
\begin{equation} \label{eq:18}
n_{EV}^{id}(T, \mu_1,\mu_2,...)=\frac{\sum_i n_i^{id}(T, \bar\mu_1,\bar\mu_2,...)}{1+\sum_k V_{ev,k}n_k^{id}(T,\bar{\mu_k})}
\end{equation}	 	
\begin{equation} \label{eq:19}
\epsilon_{EV}^{id}(T, \mu_1,\mu_2,...)=\frac{\sum_i \epsilon_i^{id}(T, \bar\mu_1,\bar\mu_2,...)}{1+\sum_k V_{ev,k}n_k^{id}(T,\bar{\mu_k})}
\end{equation}	 
\begin{equation} \label{eq:20}
s^{id}_{EV}(T, \mu_1,\mu_2,...)=\frac{\sum_i s_i^{id}(T, \bar\mu_1,\bar\mu_2,...)}{1+\sum_k V_{ev,k}n_k^{id}(T,\bar{\mu_k})}
\end{equation}	 

\begin{figure*}[htb!]
\centering
\includegraphics[scale=0.68]{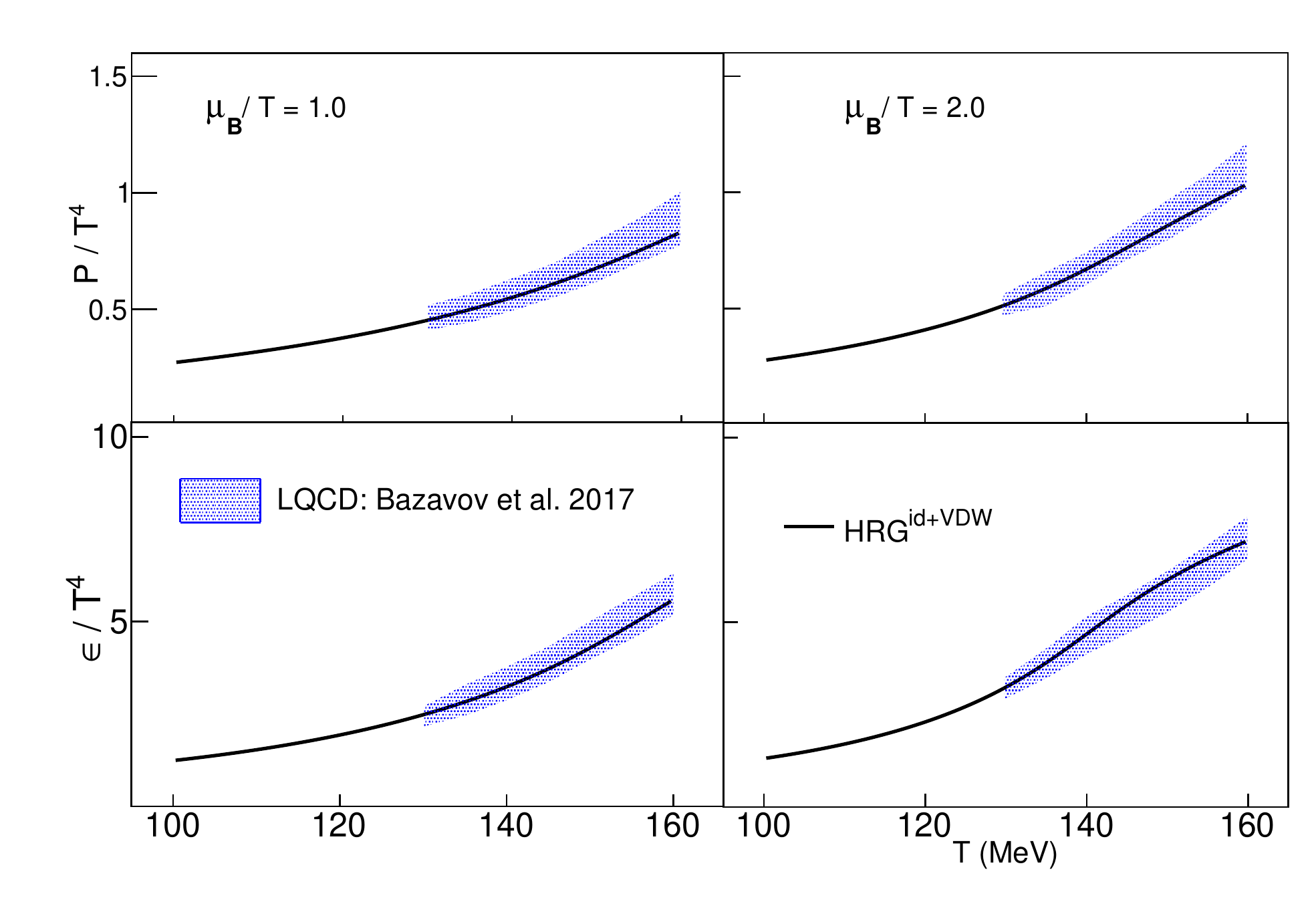}
\caption{The pressure and energy density from lattice simulation \cite {ref22} are compared with ideal HRG + VDW, as a function of $T$, with the VDW constants for interactions 
between (anti)baryons only were obtained by simultaneous fitting of the lattice results for $\mu_{B} / T$ = 1.0 and 2.0. The repulsive interaction between mesons has been effective 
by considering meson hard-core radius, $r_{M}$ = 0.2 fm.}
\label{fig:VDWPE} 
\end{figure*}	
\subsection{van der Waals Hadron Resonance Gas (VDWHRG)}

The van der Waals equation of state, in terms of the usual van der Waals constants $a$ and $b$, pressure $p(T,\mu)$ and number density $n(T,\mu)$, in the grand canonical 
ensemble of hadrons can be  written as \cite {ref26}: \\
\begin{equation}
p(T,\mu) = p^{id}(T,\bar{\mu}) - an^2(T,\mu)
\end{equation}	
\begin{equation}
n(T,\mu) = \frac{ n^{id}(T,\bar\mu)}{1 + bn^{id}(T,\bar \mu)}
\end{equation}	 	
where $\bar{\mu}$ is the modified chemical potential and is given by:\\			 
\begin{equation}
\bar{\mu} = \mu-bp(T,\mu)-abn^2(T,\mu)+2an(T,\mu)
\end{equation}	 
The other thermodynamic quantities, the entropy density, $s(T,\mu)$ and the energy density, $\epsilon(T,\mu)$ are given by:  \\
\begin{equation}
s(T,\mu) = \frac{ s^{id}(T,\bar\mu)}{1+bn^{id}(T,\bar\mu)}
\end{equation}	  			 
\begin{equation}
\epsilon(T,\mu)=\frac{ \epsilon^{id}(T,\bar\mu)}{1+bn^{id}(T,\bar\mu)}-an^2(T,\mu)
\end{equation}	 

In VDWHRG model \cite {ref25, ref26}, the VDW interactions have been considered between (anti)baryons, only, while the interactions were neglected between pairs of baryon -
antibaryon, meson-meson, and meson-(anti)baryon. In this study, while following the VDW form of interactions between (anti)baryons, the attractive and the repulsive interactions
among meson-pairs get effective through the resonances and excluded volume effect \cite {ref57, ref58}, introduced with hard-core radius of mesons $r_{M}$, respectively.\\

The pressure in the  VDWHRG model, thus, can be written as \cite {ref24, ref25, ref26}:
\begin{equation}
p(T,\mu)= 	p_M(T,\mu) + p_B(T,\mu)+	p_{\bar B}(T,\mu)
\end{equation}
where,  
\begin{equation}
p_M(T,\mu) =\sum_{i \in M} p_i^{id}(T,\bar \mu_i^M)
\end{equation}
\begin{equation}
p_B(T,\mu) =\sum_{i \in B} p_i^{id}(T,\bar \mu_i^B)
\end{equation}
\begin{equation}
p_{\bar B} (T,\mu)  =\sum_{i \in \bar B} p_i^{id}(T,\bar \mu_i^{\bar B})
\end{equation}
$\bar \mu_i^M$, $\bar \mu_i^B$ and $\bar \mu_i^{\bar B}$ are the modified chemical potential for mesons (due to EV correction), baryons and anti-baryons (due to VDW interactions), 
respectively. $p_M(T,\mu)$, $p_B(T,\mu)$ and $p_{\bar B} (T,\mu)$ are pressure of mesons, baryons and anti-baryons, respectively.
\begin{equation}
\bar{\mu}^{B,\bar B}=\mu-bp_{B,\bar B}-abn_{B,\bar B}^2+2an_{B,\bar B}
\end{equation}	 
where $n_{B}$ is the number density of baryons and $n_{\bar B}$ is that of the anti-baryons. 
The $n_{B}$ and $n_{\bar B}$ in the VDWHRG model are given by,
\begin{equation}
n_{B,\bar B}=\frac{\sum _{i \in B,(\bar B)} n_i^{id}(T,\bar \mu_i^{B,\bar B})} {1+b\sum _{i \in  B,(\bar B)}n_i^{id}(T,\bar \mu_i^{B,\bar B}) }
\end{equation}	 

Using the VDWHRG model, though we intend to study, in general, the QCD phase boundary over wide $\mu_{B}$ range, corresponding to the experiments at RHIC, LHC 
and FAIR, the focus is on as yet less understood finite $\mu_{B}$ range in the QCD phase diagram in the ($T, \mu_{B}$)-plane. We, therefore, obtain the values of the interaction 
constants between the (anti)baryons, by contrasting, simultaneously, the lattice calculated \cite {ref22} pressure and energy-density for non-zero $\mu_{B}$. We restrict the fitting 
to the lattice results for  $\mu_{B}/T$ = 1.0 and 2.0 up to the temperature, $T \sim 160$ MeV, the range of reasonable accuracy \cite {ref22} of the lattice simulation. Further, as 
has been shown in ref.~\cite{ref22}, the ideal HRG model calculations can describe the physics of the strongly interacting matter up to small values of $\mu_{B}/T$, but fail at 
large $\mu_{B} /T$ and / or $T \ge 160 MeV$.  We take $r_{M}$ = 0.2 fm from previous studies \cite {ref60, ref61} of successful description of LQCD data at $\mu_{B}$ = 0. 
Simultaneous comparison of the central values of the lattice simulations and our model calculations for pressure and energy density for different $\mu_{B}$, has been carried out 
in terms of goodness of fits, following the criterion of minimum value of $\chi^{2}$/d.o.f. The simultaneous fit results $\chi^{2}$ / d.o.f $ = 1.15 / 26$. For all our calculations 
in this work, we consider the mass table provided in reference~\cite {ref62}. The comparison of model calculations and the lattice simulations is presented in Figure~\ref{fig:VDWPE}.\\

\subsection{$\eta /s$ -the transport coefficient}
As already discussed, most of the existing approaches of estimation of $\eta /s$ for hadron gas are based on various approximations and unique formulation for calculating the 
shear viscosity for a multi-component hadron gas at varied chemical potential and temperature of interest for the hadronic phase of QCD matter is yet to emerge. However, there 
have been some studies \cite {ref35, ref46, ref47} on $\eta /s$ of hadron gas as a function of temperature and baryon chemical potential, covering a wide range in the QCD 
($T, \mu_{B}$)-plane. In this phenomenological study, we calculate $\eta$ for the VDWHRG gas following an analytical formula as adopted in reference~\cite{ref35} for calculating 
$\eta$ and $\eta /s$ for a mixture of particle species of different masses with same hard-core radius and the same mean free path of different species. In reference~\cite {ref35}, 
the $\eta /s$ of hadron gas with excluded volume effect has been calculated for $T \ge 64.3$ MeV and $\mu_{B} \le$ 800 MeV. We consider identical formulation for estimation of 
$\eta$ in the similar ($T, \mu_{B}$) ranges for the van der Waals hadron gas also. The formula for shear viscosity of a VDW hadron gas of discrete states, relating shear viscosity
coefficient to the average momentum transfer, thus can be written as:\\

\begin{equation}
\eta_{VDW}^{id}=\frac{5}{64\sqrt 8  }\sum_{i \in (M,B,\bar B)} \frac{ <|P_i|>n_i(T,\mu)}{ n(T,\mu) r_{i}^2 }
\label{eq:eta}
\end{equation}
where r is the hard-core radius of the constituents, $n_i$ is the number density of $i^{th}$ hadron, ${n(T, \mu)}$ is the number density of all the hadrons, resonances 
of the  considered van der Walls gas and $<|P_i|>$  is the average momentum given by: 	
\begin{equation}
<|P_i|>	=\frac{\int_{0}^{\infty}  \frac{p^3dp}{\exp [(\sqrt{(p^2+m_i^2)}-\bar \mu_i)/T]\pm 1} }{\int_{0}^{\infty}  \frac{p^2dp}{\exp [(\sqrt{(p^2+m_i^2)}-\bar \mu_i)/T]\pm 1} }
\end{equation} 
where, $\bar \mu_i$ is the corresponding modified chemical potential of the $i^{th}$ hadron.The (+) and (-) sign corresponds to fermions and bosons respectively.\\

The equation~\ref{eq:eta} for $\eta$ for the mixture of the gas, in the limit $T \ll m$ and $m_i = m$, reduces to that for a non-relativistic gas of hard-core sphere, having 
dependence only on temperature, radius of the hard core sphere and mass of the particles. In this work the shear viscosity for the considered VDWHRG has been calculated with:
\begin{equation}
\eta = \eta_{M} + \eta_{B} + \eta_{\bar B}
\label{eq:etaVDW} 
\end{equation}

The subscripts $M$, $B$ and $\bar B$ in equation~\ref{eq:etaVDW} stand for meson, baryon and anti-baryon, respectively. It may be noted, in the considered model for this work,
$r_{i}$ is 0.2 fm for $\eta_{M}$ and $\sim$ 0.62 fm (corresponding to interaction constant $b$ = 4.08 $fm^3$) for $\eta_{B}$ and $\eta_{\bar B}$.\\

\section{Results}		
Different parts of the wide range of $\mu_{B}$ of our study differ widely in terms of configuration of constituent hadrons and so in terms of thermodynamic variables and transport 
properties of the medium. We, therefore, present our results for two different interesting regions of low and high $\mu_{B}$. \\ 

We reiterate that all calculations in the VDWHRG model are worthy of attention only for the hadronic phase of the QCD. The VDWHRG calculations beyond a likely point of discontinuity 
or break in the monotonicity, if any, in the temperature dependent study of $s /T^3$ and $\eta /s$ for hadron gas, therefore, may be useful in presentations, only, to make the position of 
such a point more perceptible. Beyond any such point for a given $\mu_{B}$, numerical values from HRG-based calculations, without considering an EoS for partonic phase, do not 
carry actual physical properties of de-confined partonic matter. This analysis, thus, aims at constraining the locations of appearance of discontinuity or change in monotonicity of the 
observables in hadronic phase, indicating tendency towards phase transition or crossover. \\

\subsection{Low $\mu_{B}$ and high $T$ region}
\begin{figure}[htb!]
\centering
\includegraphics[scale=0.42]{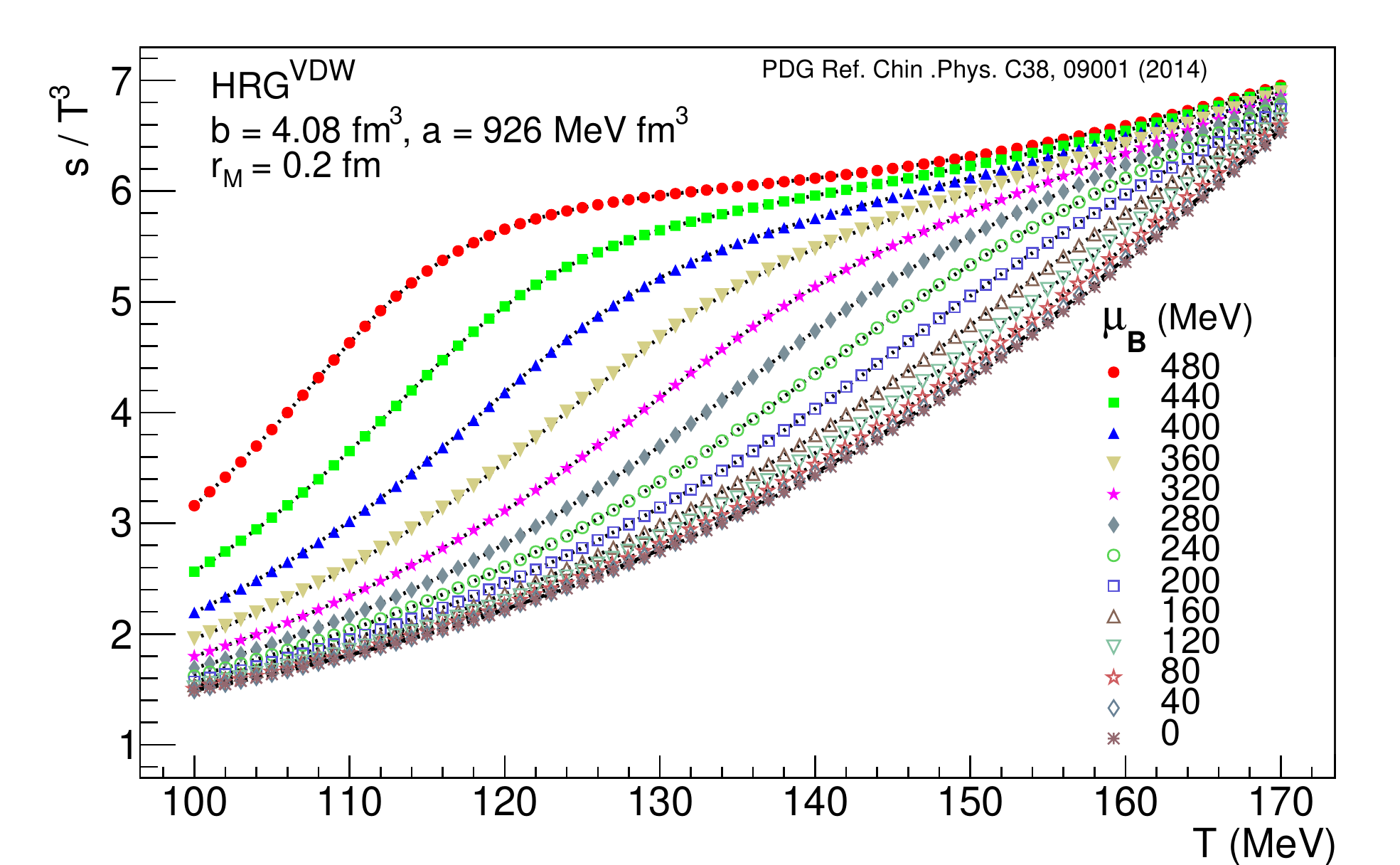}
\caption{The temperature dependent $s /T^3$ for hadron resonance gas with VDW form of interactions between (anti)baryons at different $\mu_{B}$ in the range, $\mu_{B} \sim$ 
0 to 480 MeV, covering the RHIC-BES program. The temperature dependence of the observable has been calculated at an temperature interval of 1 MeV. The dotted lines connect 
the calculated points for a given $\mu_{B}$}
\label{fig:entropy01} 
\end{figure}

\begin{figure}[htb!]
\centering
\includegraphics[scale=0.42]{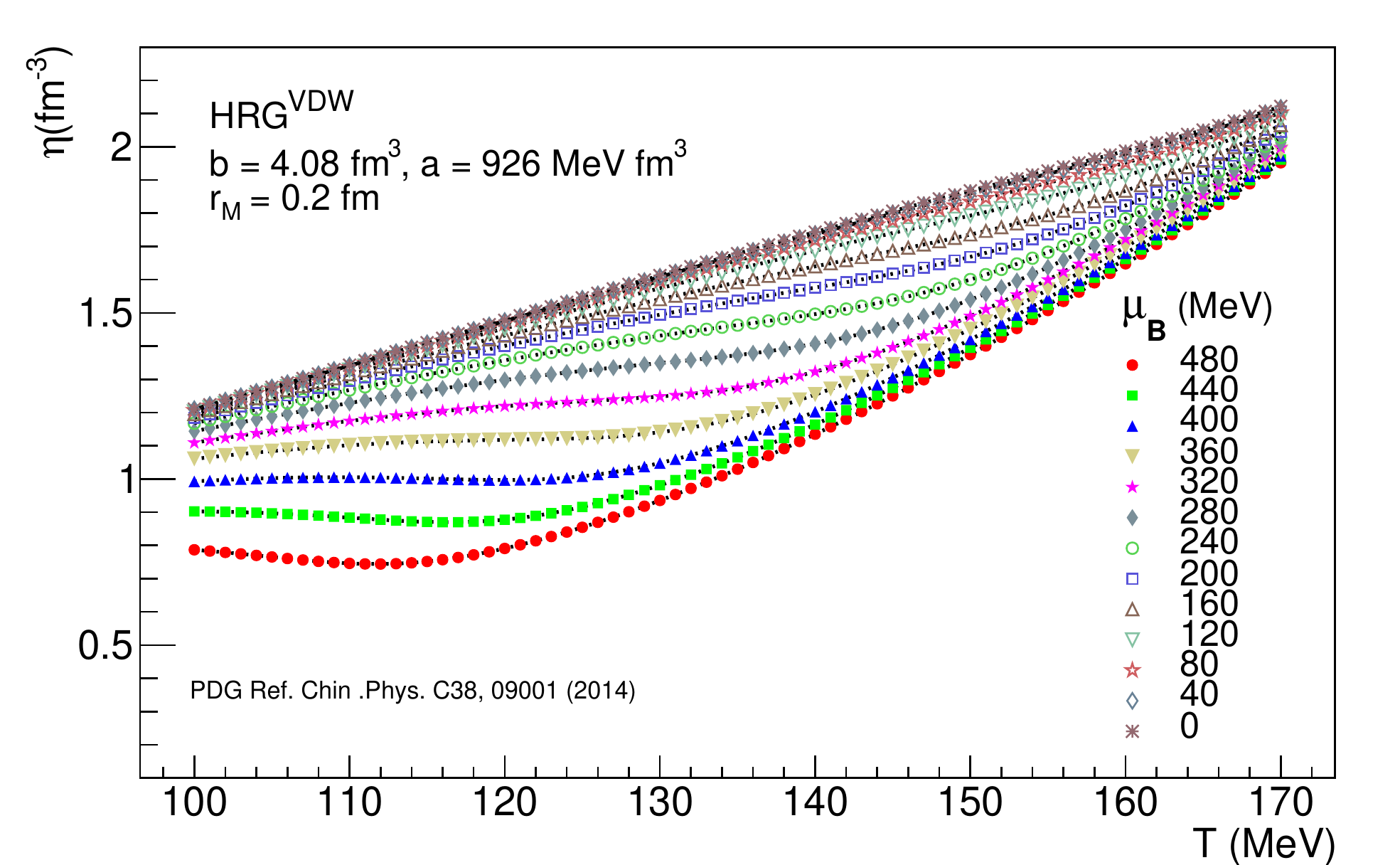}
\caption{The temperature dependent $\eta$ for system and conditions described in the caption of the figure~\ref{fig:entropy01}.}
\label{fig:eta01} 
\end{figure}

\begin{figure}[htb!]
\centering
\includegraphics[scale=0.42]{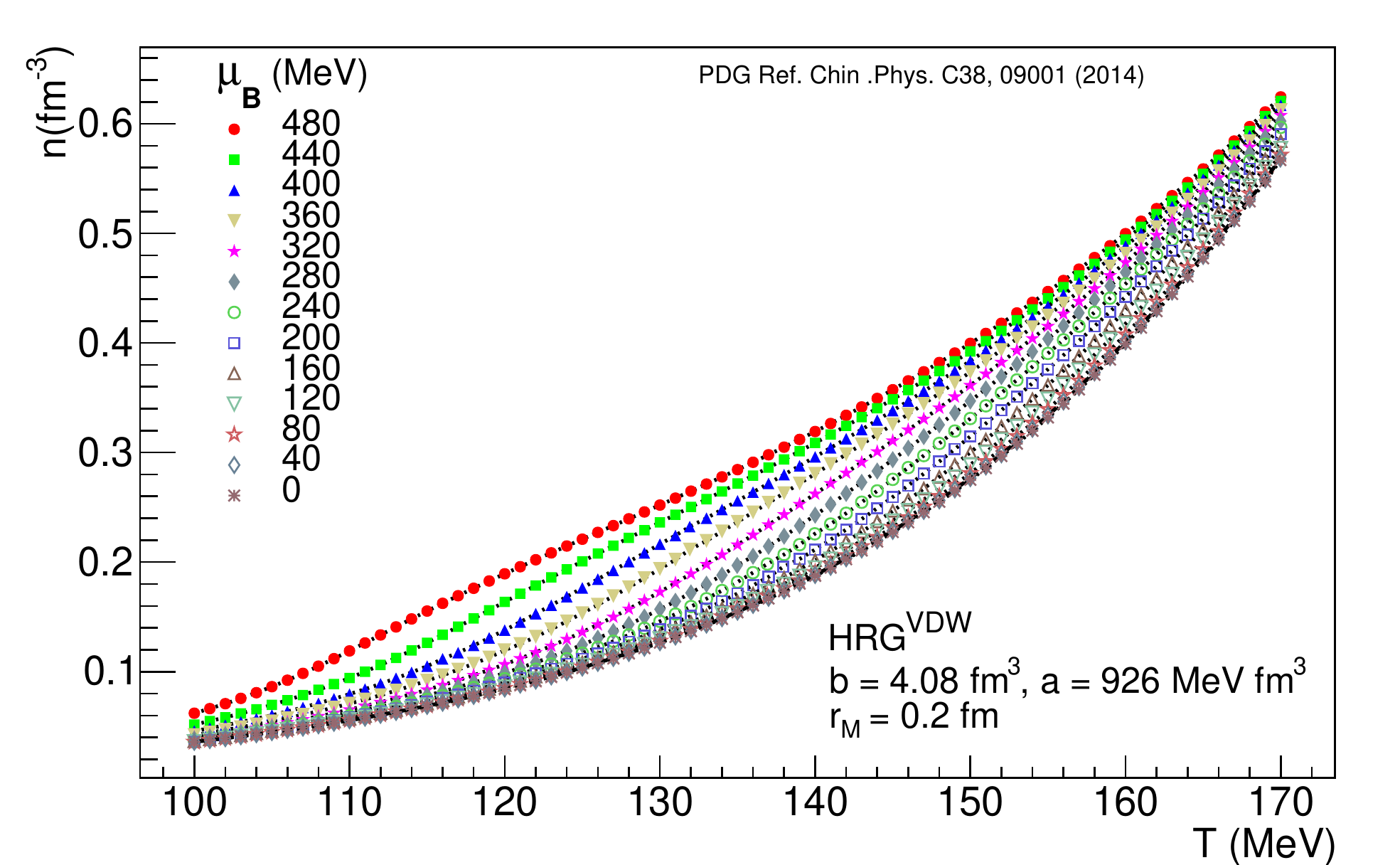}
\caption{The temperature dependent number density for system and conditions described in the caption of the figure~\ref{fig:entropy01}.}
\label{fig:number01} 
\end{figure}

\begin{figure}[htb!]
\centering
\includegraphics[scale=0.42]{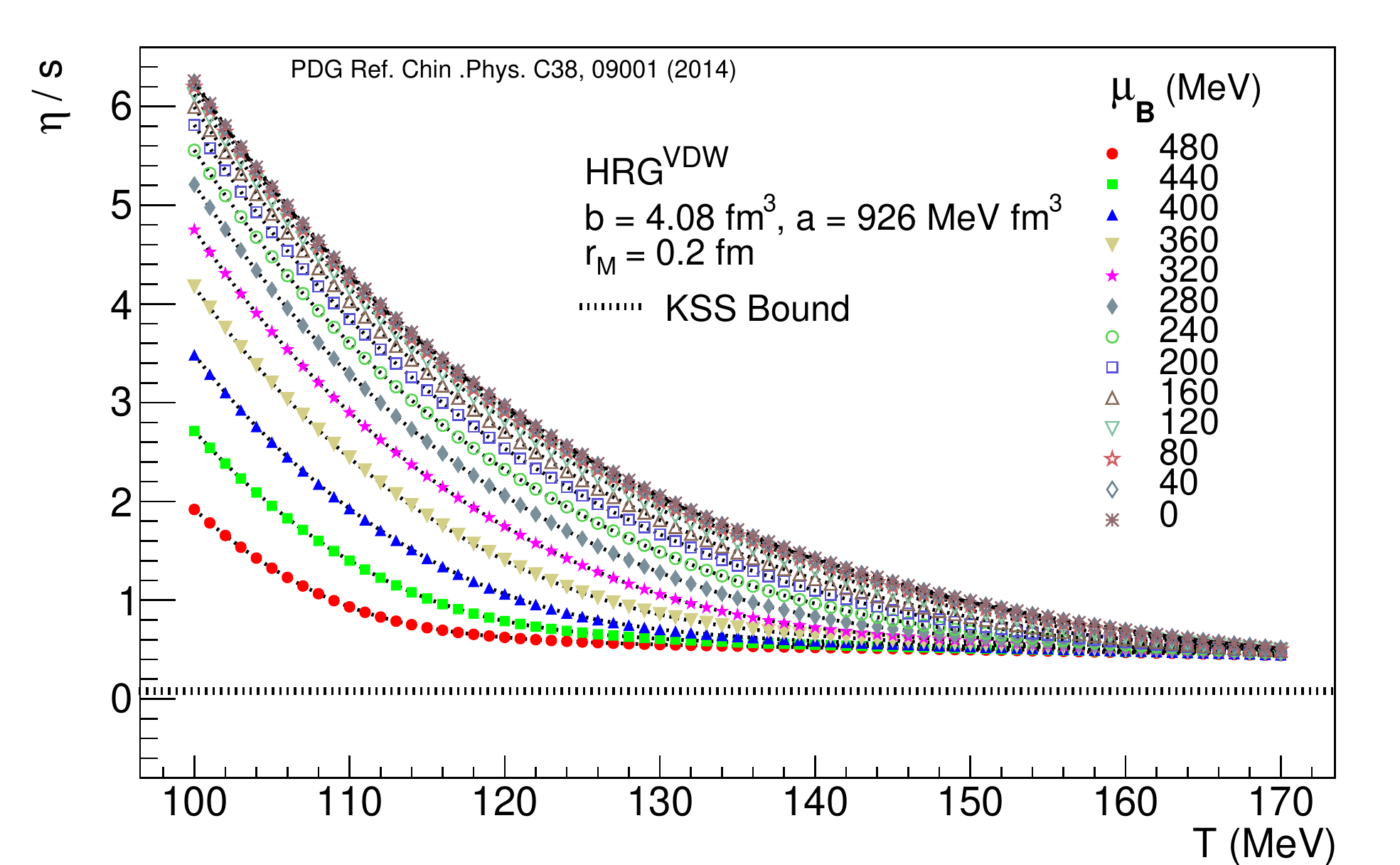}
\caption{The temperature dependent $\eta /s$ for system and conditions described in the caption of the figure~\ref{fig:entropy01}.}
\label{fig:etaBYs01} 
\end{figure}

Figures~\ref{fig:entropy01}, \ref{fig:eta01}, \ref{fig:number01} and \ref{fig:etaBYs01}, present the temperature dependent entropy density, shear viscosity, number density and 
the ratio of shear viscosity and the entropy density, respectively, for van der Waals hadron gas for $\mu_{B} \sim$ 0 to 480 MeV, corresponding to experiments at LHC and the 
first phase of RHIC-BES program. \\

Figure~\ref{fig:entropy01} shows that $s /T^3$ for zero and low $\mu_{B}$ monotonically increases with temperature. However, as the $\mu_{B}$ increases, the shape of 
temperature dependence of $s /T^3$ starts loosing monotonicity. The smooth change in shape increases with increasing $\mu_{B}$. Similar feature of smooth change in shape
with increasing $\mu_{B}$ appears in temperature dependent $\eta$ also, as can be seen in figure~\ref{fig:eta01}. In the low $\mu_{B}$ region of the hadron gas, kinetic energy of 
the constituents, dominantly the mesons, increase with temperature, resulting in to larger momentum transfer and so increase in shear viscosity. Increase in $\mu_{B}$ results in to
a denser medium with an increased relative abundance of baryons over mesons. At higher $\mu_{B}$, therefore, the effect of the VDW interactions between (anti)baryons contribute
more, reducing the momentum transfer and so the shear viscosity. The change in temperature dependent profile of entropy density and shear viscosity with increasing $\mu_{B}$ 
can be attributed to the effect of van der Waals interactions on number density as depicted in figure~\ref{fig:number01}. It is clear from figure~\ref{fig:etaBYs01}, the $\eta /s$ of the
hadron gas at a given $\mu_{B}$ decreases with temperature and reaches smoothly at a common minimum, the likely lower bound of $\eta /s$ for hadron gas in the low $\mu_{B}$-
range, as expected in the crossover region of the QCD phase boundary. Hadron Gas with higher $\mu_{B}$ reaches a minimum $\eta /s$ at lower temperature. There is no signature 
of critical point in the considered $T$ and $\mu_{B}$ ranges, in consistent with the results from the lattice calculations at finite $\mu_{B}$ \cite {ref22}, which exclude the possibility 
of having the critical point $T \gtrsim $ 135 MeV. The conjectured universal lower bound \cite{ref28, ref34} of the value of $\eta /s$ is included in figure~\ref{fig:etaBYs01}, for reference. \\

At this point, we compare values of $\eta /s$ of the hadron gas at $\mu_{B}$ = 0, in the vicinity of the crossover temperature, as estimated in several studies, following different 
methodologies. The closest match to the estimated $\eta /s$ ($\sim$ 0.69) at cross-over temperature $\sim$ 160 MeV of the present work in VDWHRG model is $\sim$ 0.62 of 
reference \cite {ref39}, where the entropy density of the hadron gas, including the Hagedorn states, has been calculated in the EVHRG model and the shear viscosity has been 
evaluated using Kubo relation. Another study in EVHRG model has been presented in reference~\cite {ref35}, where shear viscosity has been calculated for hadron gas with the 
same analytical formula that we follow in the present study. The study with different hard core radius ($r$) of constituent hadrons yields $\eta/s$ $\approx$ 0.49 and 0.29 for 
$r =$ 0.3 $fm$ and 0.5 $fm$, respectively, at $T$ = 160 MeV. The $\eta/s$ reaches $\approx$ 0.24 for $r$ $\approx$ 5.3 $fm$ at $T$ = 180 MeV. The study \cite {ref46} involving 
Kubo formalism and microscopic transport calculation within the UrQMD model yields a minimum $\eta/ s$ $\approx$ 0.9. The $\eta/ s$ calculation in Chapman Enskog and 
K-matrix formalism \cite {ref43}, for a hadronic gas of a mixture of $\pi$-$K$-$N$-$\eta$ and fifty seven resonances with masses up to 2 GeV reaches $\sim$ 0.4 at $T$ = 160 MeV. 
In reference~\cite{ref45}, the $\eta /s$ has been evaluated within the relaxation time approximation of Boltzmann equation. The relaxation time has been calculated by evaluating 
the rates of meson scatterings in a linear $\sigma$-model. It has been shown \cite{ref45} that for  sigma meson of mass $m_{\sigma}$ = 900 MeV, the minimum of $\eta /s$ 
becomes $\approx$ 0.12 (close to the KSS value, $\approx 0.08$) near the vicinity of the crossover temperature, which, however, appears at $T \sim$ 245 MeV, in the study.  \\

\subsection{High $\mu_{B}$ and low $T$ region} 
Next, we focus on the results of our study with the VDWHRG model in the high $\mu_{B}$-range for $\mu_{B} \sim$ 660 to 750 MeV, where variation in non-monotonic structure 
shows up, indicating onset of first order phase transition and crossover in our calculations of shear viscosity, entropy density, and the ratio of the two in lower temperature range. 
For a convenient presentations, the $T$-dependent profile of the observables have been calculated at an interval of 0.3 MeV and the calculated points are connected with dotted 
lines to show the discontinuity, in the profile.  \\ 
\begin{figure}[htb!]
\centering
\includegraphics[scale=0.43]{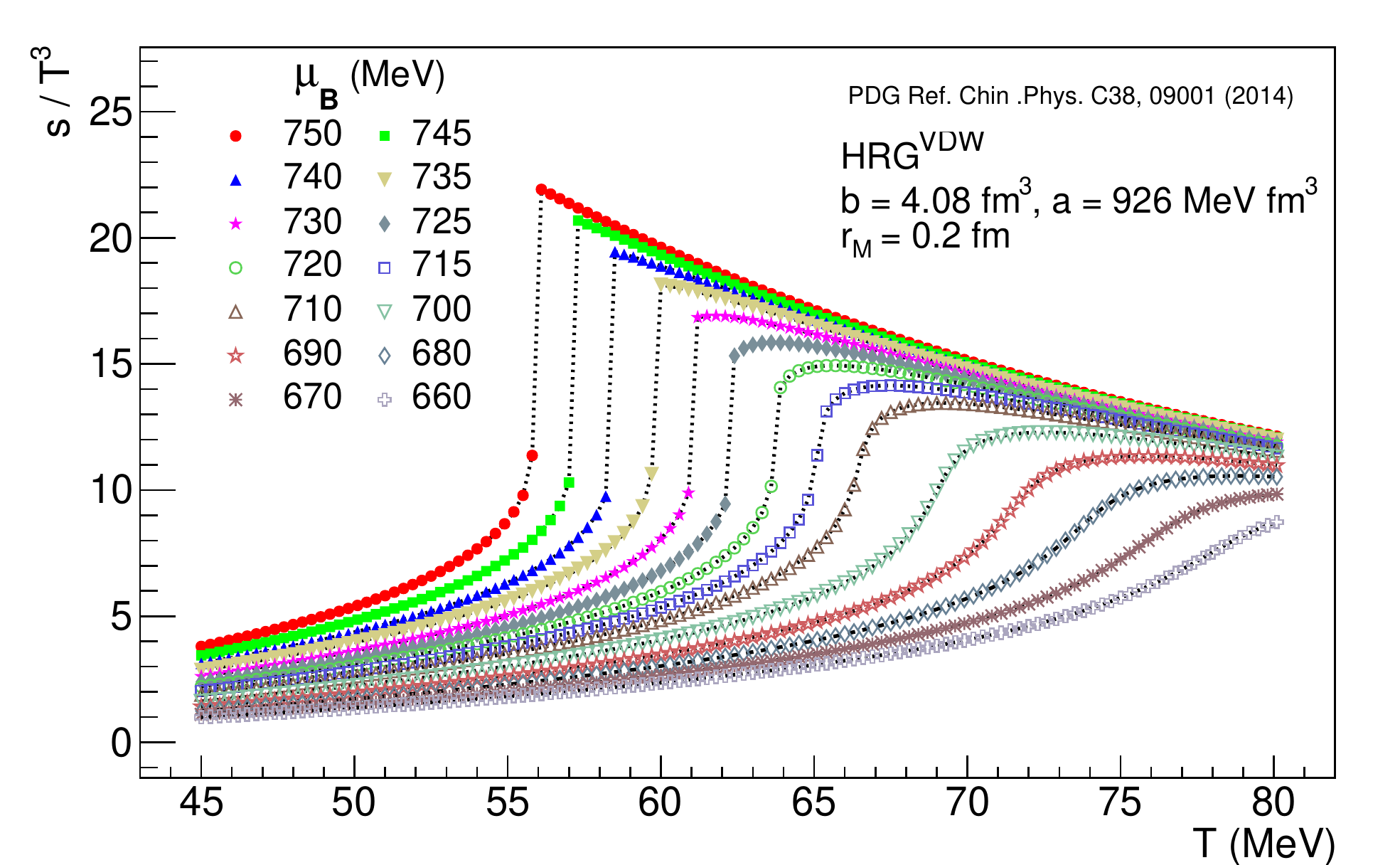}
\caption{The temperature dependent entropy density ($s$) for hadron resonance gas at $\mu_{B} \sim$ 660 to 750 MeV, the $\mu_{B}$-range where variation in non-monotonic 
structure shows up, indicating likely locations of first order phase transition and crossover along the hadronic phase boundary. The temperature dependence of the observable 
has been calculated at an temperature interval of 0.3 MeV. The dotted lines connect the calculated points for a given $\mu_{B}$.}
\label{fig:entropy} 
\end{figure}	

Figure~\ref{fig:entropy} presents temperature dependence of $s / T^3$ for the $\mu_{B}$-range of interest, at intervals. It is clear from the figure that $s / T^3$ rises fast over a 
small change in $T$. While the rise in $s / T^3$ is comparatively smooth at lower values of the considered $\mu_{B}$-range, the discontinuity in the temperature dependent profile
appears with increasing $\mu_{B}$, indicating shift from the crossover to first order phase transition or vice-versa is happening within the presented $\mu_{B}$-range. The structure 
in $T$-dependent entropy density ($s /T^3$), the thermodynamic variable, is rather expected for a gas following the van der Waals form of EoS, while the exact location of such 
structure in the ($T, \mu_{B}$)-plane depends on the values of the interaction constants.  \\  
\begin{figure}[htb!]
\centering
\includegraphics[scale=0.42]{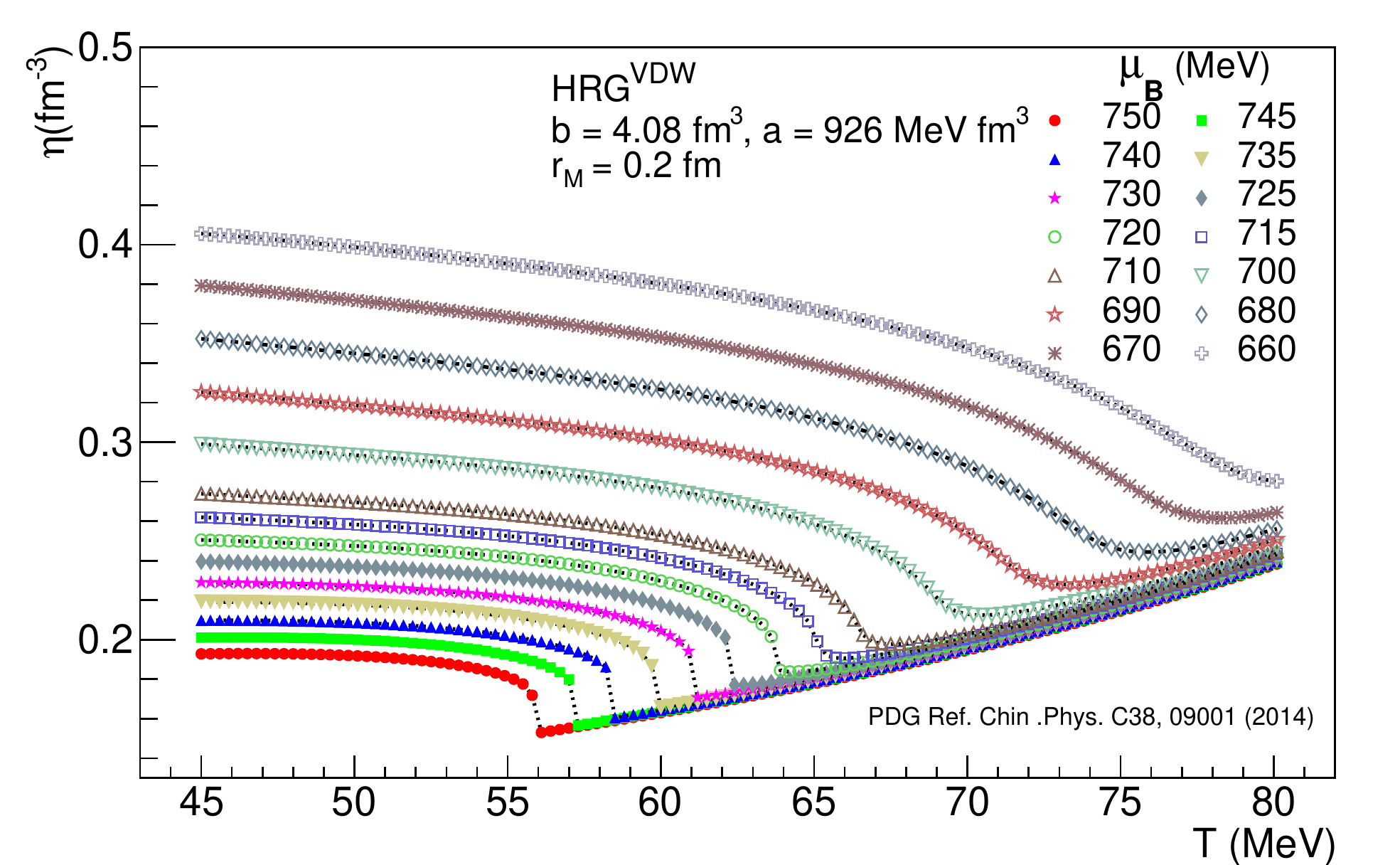}
\caption{The temperature dependent $\eta$ for system and conditions described in the caption of the figure~\ref{fig:entropy}.}
\label{fig:eta} 
\end{figure}	

\begin{figure}[htb!]
\centering
\includegraphics[scale=0.43]{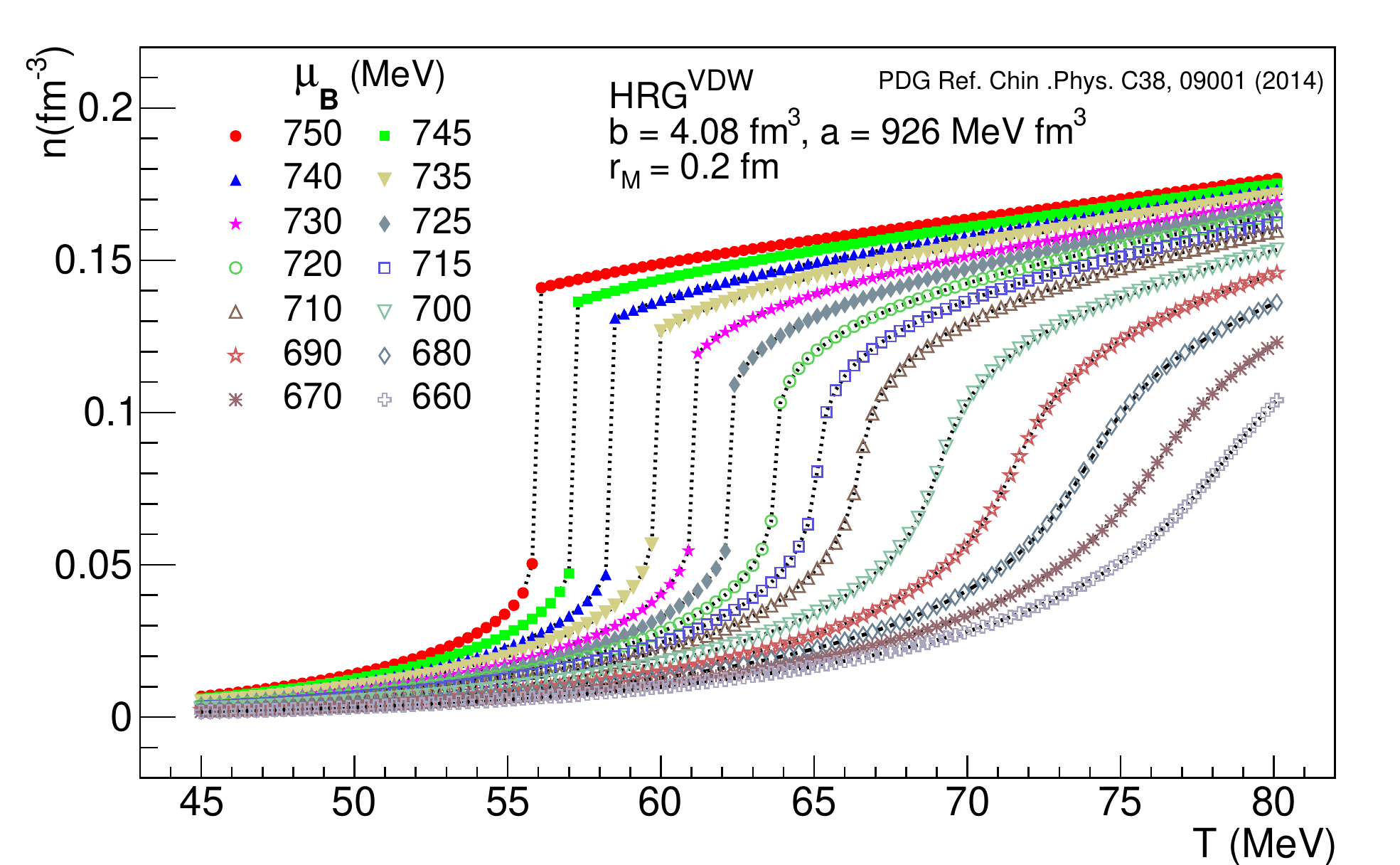}
\caption{The temperature dependent number density for system and conditions described in the caption of the figure~\ref{fig:entropy}.}
\label{fig:number} 
\end{figure}

In figure~\ref{fig:eta}, the temperature dependence of $\eta$ for the same $\mu_{B}$-range of interest and at similar regular intervals are depicted. In figure~\ref{fig:eta01},
we have seen that the temperature dependent $\eta$ at very low $\mu_{B}$ starts deviating from the monotonicity with increasing $\mu_{B}$. As seen in figure~\ref{fig:eta}, 
the temperature dependent $\eta$ at much higher $\mu_{B}$ and at lower temperature exhibit more prominent deviation. In this high $\mu_{B}$ and low temperature region,
the shear viscosity decreases with temperature as the transfer of momentum decreases with stronger interactions among densely populated baryons. Discontinuity in $\eta$ 
takes place in this region of high $\mu_{B}$ and low temperature. The non-monotonic feature of the temperature dependent $s /T^3$ and $\eta$ in the high baryon density
region can be attributed to the temperature dependence of number density in this region, that is depicted in figure~\ref{fig:number}. \\
\begin{figure}[htb!]
\centering
\includegraphics[scale=0.42]{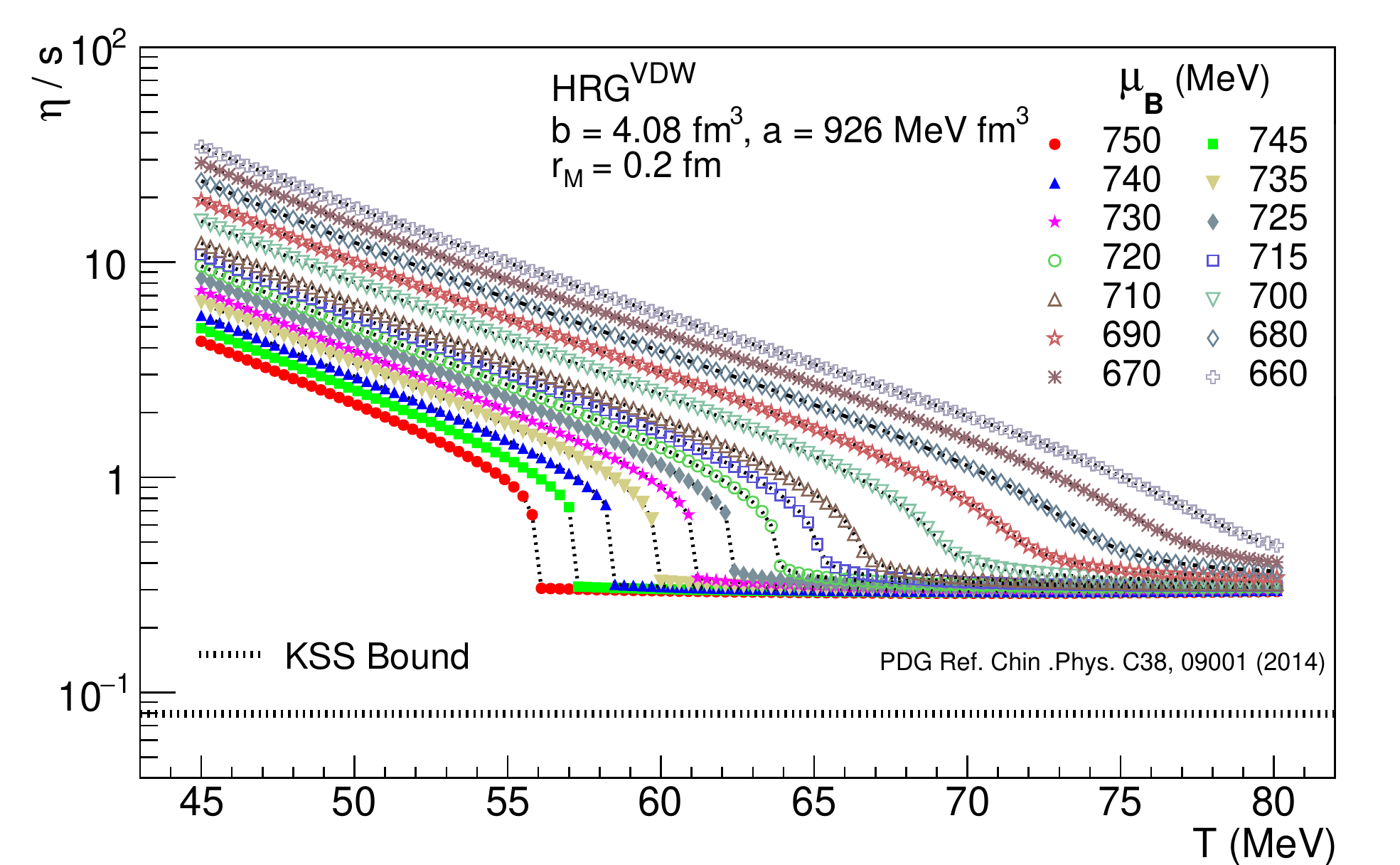}
\caption{The temperature dependent $\eta /s$ for system and conditions described in the caption of the figure~\ref{fig:entropy}.}
\label{fig:etaBYs02} 
\end{figure}	

In figure~\ref{fig:etaBYs02}, we plot temperature dependent ratio, $\eta /s$ for the van der Waals hadron resonance gas at different $\mu_{B}$ for which the temperature dependence 
of $\eta$ and $s / T^3$ have been studied separately. The conjectured universal lower bound \cite{ref34} of $\eta /s$ has been shown in the figure. The structure in the temperature
dependent $\eta /s$ reiterates the region of first order phase transitions and crossover for the given set of values of the interaction constants. It may be noted that the estimated value 
of $\eta /s$ for the hadron gas does not reach the KSS bound, in consistent with most of the previous studies on $\eta /s$ for hadron gas in different models, as already discussed.
Also, at or near the probable location of the critical point, the so-called minimum of $\eta /s$ is not reached for any of the $\mu_{B}$, studied here. This feature corroborates the
conclusion of reference~\cite {ref46} that the minimum of $\eta /s$ for the QCD matter reaches at the de-confined phase rather than at the confined, hadronic phase.\\
\begin{figure}[htb!]
\centering
\includegraphics[scale=0.42]{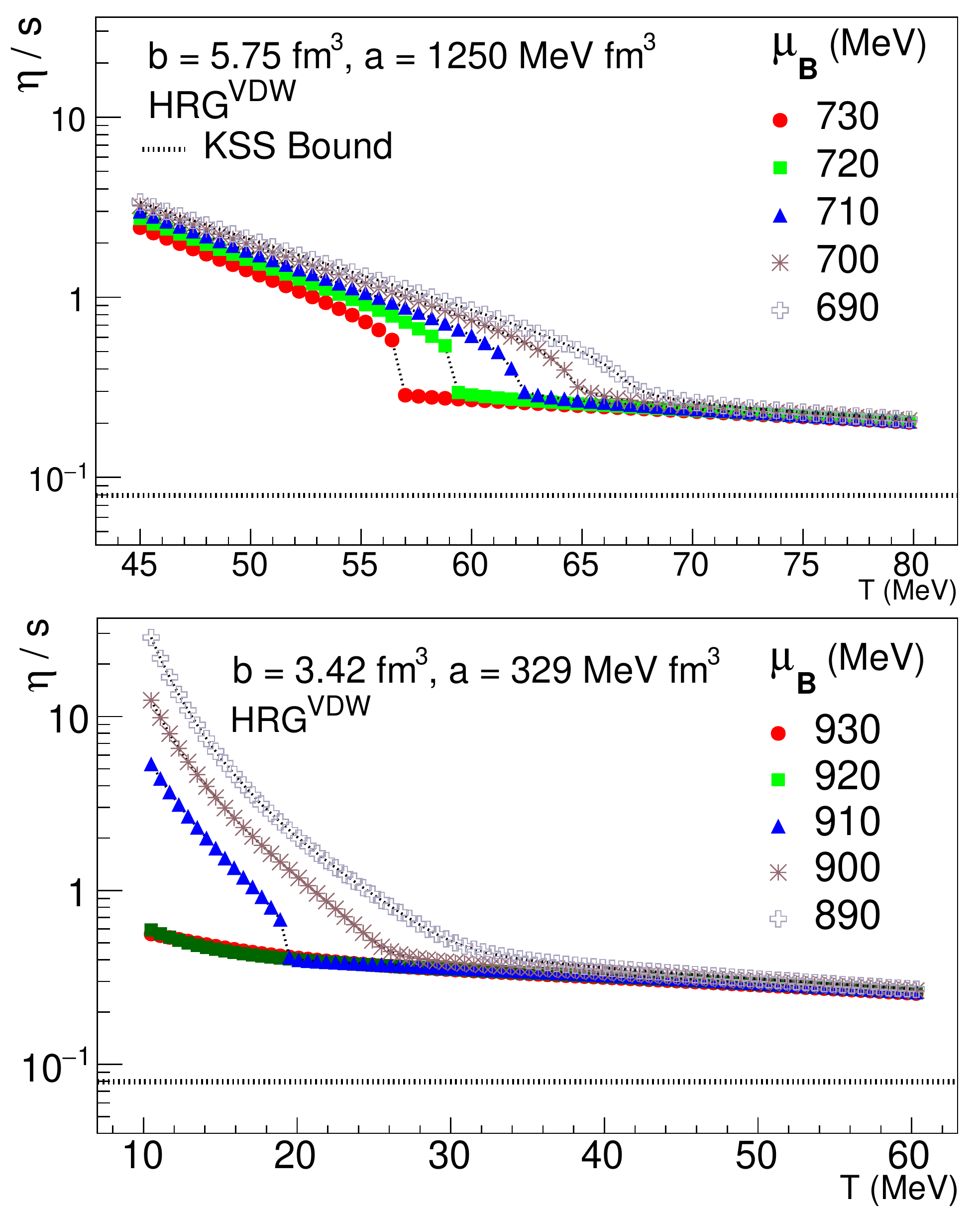}
\caption{Temperature dependent $\eta /s$ for VDW hadron gas with interaction parameters $a$ = 329 MeV $fm^3$ and $b$ = 3.42 $fm^3$ (lower panel) as used in 
reference~\cite{ref25} and $a$ =1250 MeV $fm^3$ and $b$ = 5.75 $fm^3$ (upper panel) as used in reference~\cite {ref27}.}
\label{fig:etaBYs03} 
\end{figure}	

Occurrence of similar feature of temperature dependent $\eta /s$ for VDWHRG with previously studied combinations of interaction constants is depicted in figure~\ref{fig:etaBYs03}. 
The lower panel of figure~\ref{fig:etaBYs03} shows that for the values of the constants, $a$ = 329 MeV $fm^3$ and $b$ = 3.42 $fm^3$ for ground state nuclear matter \cite {ref24, 
ref25}, the point in between regions of the first order phase transition and the crossover, in terms of discontinuity in $\eta /s$, appears near $T \simeq$ 19.5 MeV and at $\mu_{B} 
\simeq$ 910 MeV (we calculate $\eta /s$ at intervals of 10 MeV in $\mu_{B}$ and 0.3 MeV in $T$, while in references~\cite {ref24} and \cite{ref25}, the critical point at the end of 
nuclear liquid-gas phase transition has been shown to be at $T$ = 19.7 MeV and $\mu_{B}$ = 908 MeV. In the upper panel of figure~\ref{fig:etaBYs03}, $a$ =1250 $\pm$ 150 MeV
$fm^3$ and $r$ = 0.7 $\pm$ 0.05 $fm$ (corresponding to $b \sim$ 5.75 $fm^3$), as used in estimating \cite {ref27} QCD critical point, the $T$-dependent $\eta /s$ indicates the point 
between the first order phase transition and the crossover resides at $T \simeq$ 62.1 MeV and $\mu_{B} \simeq$ 720 MeV against the estimated critical point at $T$ = 62.1 MeV 
and $\mu_{B}$ = 708 MeV (where $\partial p/ \partial n$ equals to zero) in reference~\cite {ref27}. The striking similarities in results on the appearance of the point between regions
of the first order phase transition and the crossover, in terms of discontinuity in $\eta /s$, coinciding with the estimated location of critical point in the previous studies for respective
combinations of interaction constants, corroborate our consideration of the onset of phase transition and crossover in hadronic phase being indicative to probable location of critical 
point. In terms of discontinuity in $s /T^3$ and $\eta /s$ in the hadronic phase in a given $\mu_{B}$, the probable location of the QCD critical point is estimated to lie around around 
$T \sim$ 65 MeV and $\mu_{B} \sim$ 715 MeV.\\  

\section{Discussions and summary}
We study the temperature dependence of $s / T^3$ and $\eta /s$ for hadron gas with van der Waals form of EoS, satisfying lattice EoS for the strongly interacting QCD matter. 
The study helps constraining the region of onset of the first order phase transition and crossover in the hadronic phase of the QCD matter, eventually indicating the probable location 
of the critical point.  \\

At vanishing or low $\mu_{B}$, the number density of meson-dominated hadron gas increases with increasing temperature and the van der Waals form of interactions between 
(anti)baryons do not influence much the behaviour of $s /T^3$ or $\eta /s$. On the other hand, in the large $\mu_{B}$-region, where the system becomes more and more 
baryon-rich with increasing $\mu_{B}$, the effects of interactions between (anti)baryons on $s /T^3$ or $\eta /s$ start appearing. Stronger interactions in the region 
of low $T$ and high $\mu_{B}$ results into non-trivial structure in temperature dependent $s /T^3$ or $\eta /s$. \\

In this phenomenological work, $\eta$ for hadron gas has been calculated from an approximate formulation in molecular kinetic theory. On the basis of our calculations, only, 
it will not be proper to conclude on quantitative estimations of $\eta /s$ for the hadron gas. But that does not cause impediment to identification of regions of interest in phase 
diagram through qualitative comparisons of non-trivial structures, particularly when the revelation is corroborated with complementary study in terms of $s/T^3$. Also, the 
approximate formula does not cause any hindrance in finding the location of end point of discontinuity, appearing for a given $\mu_{B}$ in temperature dependent study of 
$\eta /s$, that matches with the location of critical point estimated in other studies in terms of different observables.    \\  

The van der Waals form of EoS, by way of its construction, reveals the first order phase transition. Discontinuities or non-monotonic structure in temperature dependent $s /T^3$, 
$\eta$ and $\eta /s$ for varied $\mu_{B}$ reveal the predicted signature of first order phase transitions or crossovers over certain ranges of $\mu_{B}$, that is determined by the 
values of the van der Waals interaction constants. Reliability of predicted regions of phase transition or crossover and so the probable location of critical point depends largely on 
choice of the interaction constants. We  obtain the constants by simultaneous fit of lattice calculations of thermodynamic variables at different finite $\mu_{B}$ of QCD matter. We 
find the possible location of the QCD critical point lies around $T \sim$ 65 MeV and $\mu_{B} \sim$ 715 MeV.\\ 

Interestingly, the probable location of the critical point, estimated with the interaction constants obtained by comparing lattice EoS for low but finite $\mu_{B}$, in this study, is not very 
different from that found (at $T$ = 62.1 MeV and $\mu_{B}$ = 708 MeV, where $\partial p/ \partial n$ equals to zero) in reference~\cite {ref27} by fitting the LQCD EoS for $\mu_{B}$ 
= 0. This shows a weak $\mu_{B}$-dependence of the VDW constants, obtained by matching lattice EoS within the range of $\mu_{B}$ from 0 to $\sim$ 300 MeV, in estimation of 
probable QCD critical point from analysis of VDWHRG. In view of this observation, in the absence of an EoS for QCD matter from first principle calculation for higher $\mu_{B}$, 
our approximation on validity of VDW constants obtained for low $\mu_{B}$ in the high $\mu_{B}$ region appears reasonable.\\

We recollect at this stage that the lattice calculation \cite {ref22} at finite $\mu_{B}$ excludes the possibility of having the critical point in the ranges $T$ ($\gtrsim 135$ MeV) and 
$\mu_{B}$ ($\lesssim 300$ MeV). Also, analysis of RHIC heavy-ion data in Finite Size Scaling \cite {ref63} method rules out a possible location of critical point below $\mu_{B}$ = 
400 MeV. The location of the QCD critical point estimated by this work and the work presented in reference~\cite {ref27}, in the VDWHRG model, is comparable with that obtained 
by holographic Einstein-Maxwell-Dilaton (EMD) model \cite {ref64, ref65}. The EMD model calculations, performed in classical limit of the gauge / string duality, are successfully
applicable along the QCD phase boundary, the strongly coupled infrared regime of the QCD. The QGP EoS in the EMD model, that agrees well with the lattice results \cite {ref22} 
at finite temperature and baryon density, finds a critical point at $T$ = 89 MeV and $\mu_{B}$ = 724 MeV. In summary, this study, along with the complementing results from other 
recent studies, thus indicates that the QCD critical point may probably be found in the baryon-rich matter, corresponding to $\mu_{B} \gtrsim $ 700 MeV, likely to be formed in future
experiments of heavy-ion collisions at RHIC BES-II, CBM, NICA or J-PARC-HI. \\


\begin{thebibliography}{99}

\bibitem{ref01} J. C. Collins and M. J. Perry, Phys. Rev. Lett. {\bf 34}, 1353 (1975).
\bibitem{ref02} E. Shuryak, Phys. Rep. {\bf61}, 71 (1980).
\bibitem{ref03} I. Arsene et al., BRAHMS Collaboration, Nucl. Phys. {\bf A757}, 1 (2005).
\bibitem{ref04} B. B. Back et al., PHOBOS Collaboration, Nucl. Phys.  {\bf A757}, 28 (2005).
\bibitem{ref05} J. Adams et al., STAR Collaboration, Nucl. Phys. {\bf A757}, 102 (2005).
\bibitem{ref06} K. Adcox et al., PHENIX Collaboration, Nucl. Phys. {\bf A757}, 184 (2005).
\bibitem{ref07} B. Muller, J. Schukraft, and B. Wyslouch, Annu. Rev. Nucl. Part. Sci. {\bf 62}, 361 (2012).
\bibitem{ref08} K. G. Wilson, Phys. Rev. {\bf D10}, 2445 (1974).	
\bibitem{ref09}  F. Karsch, K. Redlich, and A. Tawfik, Phys. Lett. {\bf B571}, 67 (2003).
\bibitem{ref10} A. Andronic, P. Braun-Munzinger, J. Stachel, and M. Winn, Phys. Lett. {\bf B718}, 80 (2012).
\bibitem{ref11} M. Albright, J. Kapusta, and C. Young, Phys. Rev. {\bf C90}, 024915 (2014).
\bibitem{ref12} X. Luo and N. Xu, Nucl. Sci. Tech., {\bf 28}, 112 (2017).
\bibitem{ref13} Z. Fodor and S. D. Katz JHEP {\bf 04} 050 (2004). 
\bibitem{ref14} Y. Aoki, G. Endrodi, Z. Fodor, S. D. Katz and K. K. Szabo, Nature {\bf 443}, 675 (2006).
\bibitem{ref15} M. Asakawa and K. Yazaki, Nucl. Phys. {\bf A504} 668 (1989).
\bibitem{ref16} F. Karsch and M. Lutgemeier, Nucl. Phys. {\bf B550}, 449 (1999).
\bibitem{ref17} M. Halasz, A. Jackson, R. Shrock, M. A. Stephanov, and J. Verbaarschot, Phys. Rev. {\bf D58}, 096007 (1998).
\bibitem {ref18} STAR Note 0598: BES-II whitepaper: http : //drupal.star.bnl.gov/ST AR/starnotes/public/sn0598.
\bibitem {ref19} T. Ablyazimov et al., CBM Collaboration, Eur. Phys. J. {\bf A53}, 60 (2017).
\bibitem{ref20} F. Becattini and R. Stock, Eur. Phys. J. {\bf A52}, 234 (2016).
\bibitem{ref21} H. Sako et al., Nucl. Phys. {\bf A956}, 850 (2016).
\bibitem{ref22} A. Bazavov, et al., Phys. Rev. {\bf D95} 054504 (2017).
\bibitem{ref23} V. Vovchenko, D. V. Anchishkin, and M. I. Gorenstein, J. Phys. {\bf A48}, 305001 (2015).
\bibitem{ref24} V. Vovchenko, D. V. Anchishkinm, and M. I. Gorenstein, Phys. Rev. {\bf C91}, 064314 (2015).
\bibitem{ref25} V. Vovchenko, D. V. Anchishkin, M. I. Gorenstein and R. V. Poberezhnyuk, Phys. Rev. {\bf C92}, 054901 (2015).
\bibitem{ref26} V. Vovchenko, D. V. Anchishkinm, and M. I. Gorenstein, Phys. Rev. Lett {\bf 118}, 182301 (2017).
\bibitem{ref27} S. Samanta and B. Mohanty, Phys. Rev. {\bf C97}, 015201 (2018). 
\bibitem{ref28} L.P. Csernai, J.I. Kapusta, and L.D. McLerran, Phys. Rev. Lett. {\bf 97} 152303 (2006).
\bibitem{ref29} R. A. Lacey, et al. Phys. Rev. Lett. {\bf 98}, 092301 (2007).
\bibitem{ref30} J. Chen, Y. Li, Y. Liu, and E. Nakano, Phys. Rev. {\bf D76}, 114011 (2007).
\bibitem{ref31} J. Chen, E. Nakano, Phys. Lett. {\bf B647}, 371 (2007).
\bibitem{ref32} A. Dobado, F. J. Llanes-Estrada, and J. M. Torres-Rincon, Phys. Rev. {\bf D79}, 014002 (2009).
\bibitem{ref33} A. Dobado, F. J. Llanes-Estrada, and J. M. Torres-Rincon, Phys. Rev. {\bf D80}, 114015 (2009).
\bibitem{ref34} P. Kovtun, D. T. Son, and A. O. Starinets, Phys. Rev. Lett. {\bf 94}, 111601 (2005).
\bibitem{ref35} M. Gorenstein, M. Hauer, O. Moroz, Phys. Rev. {\bf C77}, 024911 (2008).
\bibitem{ref36} J. Noronha-Hostler, J. Noronha and C. Greiner, Phys. Rev. Lett. {\bf 103}, 172302 (2009).
\bibitem{ref37} J. Noronha-Hostler, J. Noronha and C. Greiner , Phys. Rev. {\bf C86}, 024913 (2012).
\bibitem{ref38} G. P. Kadam and H. Mishra  Phys. Rev. {\bf C93}, 025205 (2016).
\bibitem{ref39} S. Pal, Phys. Lett. {\bf B684}, 211 (2010).
\bibitem{ref40} R. Kubo, Rep. Prog. Phys. {\bf 29}, 255  (1966).
\bibitem{ref41} W. A. Van Leeuwen, P. H. Polak, S. R. De Groot, Physica {\bf 63}, 65 (1973).
\bibitem{ref42} S. Plumari, A. Puglisi, F. Scardina and V. Greco, Phys. Rev. {\bf C86}, 054902 (2012). 
\bibitem{ref43} A. Wiranata, V. Koch, M. Prakash and X. N. Wang, Phys. Rev. {\bf C88}, 044917 (2013).
\bibitem{ref44} S. Gavin, Nucl. Phys. {\bf A435}, (1985) 826. 
\bibitem{ref45} P. Chakraborty and J. I. Kapusta, Phys. Rev. {\bf C83}, 014906 (2011).
\bibitem{ref46} N. Demir and S. A. Bass, Phys. Rev. Lett. {\bf 102}, 172302 (2009).
\bibitem{ref47} J.-B. Rose, J. M. Torres-Rincon, A. Schafer, D. R. Oliinychenko, and H. Petersen, arXiv:1709.03826 [nucl-th] (2017). 
\bibitem{ref48} S.Pratt, A. Baez and J. Kim, Phys. Rev. {\bf C95}, 024901 (2017).
\bibitem{ref49} P. Romatschke and U. Romatschke, Phys. Rev. Lett. {\bf 99}, 172301 (2007).
\bibitem{ref50} A. Bazavov, H.-T. Ding, P. Hegde, O. Kaczmarek, F. Karsch, E. Laermann, Y. Maezawa, S. Mukherjee et al., Phys. Rev. Lett. {\bf 113}, 072001 (2014).
\bibitem{ref51} S. Borsanyi, Z. Fodor, S. D. Katz, S. Krieg, C. Ratti, and K. Szabo, J. High Energy Phys. {\bf 01}, 138 (2012).
\bibitem{ref52} A. Bazavov et al. (HotQCD Collaboration), Phys. Rev. {\bf D86}, 034509 (2012).
\bibitem{ref53} S. Borsanyi, Z. Fodor, C. Hoelbling, S. D. Katz, S. Krieg, and K. K. Szabo, Phys. Lett. {\bf B730}, 99 (2014).
\bibitem{ref54} R. Hagedorn and J. Rafelski, Phys. Lett. {\bf B97}, 136 (1980).
\bibitem{ref55} R. Hagedorn, Z. Phys {\bf C17}, 265 (1983).
\bibitem{ref56} M. I. Gorenstein, V. K. Petrov and G. M. Zinovjev Phys. Lett. {\bf B106}, 327 (1981).
\bibitem{ref57} J. I. Kapusta and K. A. Olive, Nucl. Phys. {\bf A408}, 478 (1983).
\bibitem{ref58} D. H. Rischke, M. I. Gorenstein, H. Stoecker and W. Greiner Z. Phys. {\bf C51}, 485 (1991).
\bibitem{ref59} J. Cleymans and H. Satz, Z. Phys. {\bf C57}, 135 (1993).
\bibitem{ref60} V. Vovchenko, D. V. Anchishkinm, and M. I. Gorenstein, Phys. Rev. {\bf C91}, 024905 (2015).
\bibitem{ref61} N. Sarkar and P. Ghosh, Phys. Rev. {\bf C96}, 044901 (2017).
\bibitem{ref62} K. A. Olive et al., Chin. Phys. {\bf C38}, 090001 (2014).
\bibitem{ref63} E. S. Fraga, L. F. Palhares and P. Sorensen, Phys. Rev. {\bf C84}, 011903(R) (2011).  
\bibitem{ref64} R. Rougemont, R. Critelli, J. Noronha-Hostler, J. Noronha and C. Ratti, Phys. Rev. {\bf D96}, 014032 (2017).
\bibitem{ref65} R. Critelli, J. Noronha, J. Noronha-Hostler, I. Portillo, C. Ratti and R. Rougemont,  Phys. Rev. {\bf D96}, 096026 (2017).
\end{thebibliography}
\end{document}